# Prediction method of cigarette draw resistance based on correlation analysis

Linsheng Chen[a], Zhonghua Yu[a,*], Bo Zhang[b], Qiang Zhu[b], Hu Fan[b], Yucan Qiu[b]

[a] College of Mechanical Engineering, Zhejiang University, Hangzhou 310030, China
[b] Technology Center of China Tobacco Zhejiang Industrial Co. Ltd, Hangzhou 310008, China

Abstract：

The cigarette draw resistance monitoring method is incomplete and single, and the lacks correlation analysis and preventive modeling, resulting in substandard cigarettes in the market. To address this problem without increasing the hardware cost, in this paper, multi-indicator correlation analysis is used to predict cigarette draw resistance. First, the monitoring process of draw resistance is analyzed based on the existing quality control framework, and optimization ideas are proposed. In addition, for the three production units, the cut tobacco supply (VE), the tobacco rolling (SE), and the cigarette-forming (MAX), direct and potential factors associated with draw resistance are explored, based on the linear and non-linear correlation analysis. Then, the correlates of draw resistance are used as inputs for the machine learning model, and the predicted values of draw resistance are used as outputs. Finally, this research also innovatively verifies the practical application value of draw resistance prediction: the distribution characteristics of substandard cigarettes are analyzed based on the prediction results, the time interval of substandard cigarettes being produced is determined, the probability model of substandard cigarettes being sampled is derived, and the reliability of the prediction result is further verified by the example. The results show that the prediction model based on correlation analysis has good performance in three months of actual production.

Keywords: Agricultural products, Cigarette rolling, draw resistance monitoring, correlation analysis, machine learning.

1. Introduction

A new digital era has been opened by national policies such as "Advanced Manufacturing Leadership Strategy" in the United States, "National Industrial Strategy 2030" in Germany, "Robot Strategy" in Japan, "The future of manufacturing: a new era of opportunity and challenge for the UK" in the UK, and "Made in China 2025" in China. In this context, the plan for establishing modernized factories such as digitalization and intelligence is proposed by the China Tobacco Bureau, and its cigarette manufacturers responded positively and devoted themselves to the construction of enterprise digitalization, especially in the development and application of ERP, MES, CAD/CAM/CAT, and PDM systems[1, 2], which created a good foundation for the subsequent intelligent construction.

At present, China Tobacco is equipped with a large number of sensors and quality monitoring equipment in production, collecting more than 20T of time-series data every day, but there is a problem of insufficient depth and breadth of data mining, resulting in the potential value of massive data not being fully explored[3]. Especially in the quality control for draw resistance, the online monitoring only relies on the Cigarette Inspection System-CIS, and the finished product inspection stays at the primary stage of random sampling [4]. To enrich the quality monitoring methods, and reduce the number of substandard cigarettes missed by CIS, data-driven research on draw resistance quality improvement without additional sensors needs to be carried out.

At home and abroad, there are more academic studies and fewer systematic applications in the field of digital quality control in the tobacco manufacturing industry, so the development is in the primary stage. The existing quality control of draw resistance mainly adopts statistical indicators[5], which have problems of high dispersion, time series fragmentation and early warning lag. The application value of data has not been explored by the traditional cigarette quality control framework, such as poor model adaptiveness[6], lack of quality assessment of data [7], and no correlation analysis between multiple indicators[8]. Therefore, scholars have attempted to study the cigarette production process and analyze the large amount of collected data.

In the processing of cigarettes, quality control methods for the tobacco leaves, cut tobacco, and cigarette rolling have been studied by scholars. Cigarette leaves-oriented studies have focused on the quality of raw materials[9-11], as well as on physicochemical characterization[12-14].

Research on cut tobacco is mainly focused on the physicochemical analysis [15, 16]. Using image processing technology, Condorf[15] investigated the mapping relationship between temperature, air humidity, tobacco weight loss and tobacco images, to achieve real-time monitoring of tobacco weight during the roasting process. Martinez[17] used artificial neural networks to build a nonlinear fitting system for temperature and humidity, to predict the humidity during the drying process of tobacco leaves.

Cigarette rolling is the last and most critical process in cigarette production, a large number of monitoring technologies are used to assess the final quality of the cigarettes. Yin[18] proposed a new section geometry parameter measurement method based on machine vision technology, to measure the depth, number of grooves, and area ratio for cigarette filter rods. Cao[19] completed online measurement of cigarette circumference value based on industrial camera and vision information.

In conclusion, the above method is only for quality control of each section separately. In the actual production, the in-process products passed to the subsequent section all meet the production requirements, but there are still cigarettes that are rejected because of substandard quality in the final rolling section, indicating that the quality control technology of cigarettes still needs to be improved.

Cigarette draw resistance directly affects the taste of consumers, too much draw resistance leads to difficulty in smoking, and too little draw resistance leads to insufficient aroma, so draw resistance is one of the key quality indicators for control in cigarette rolling. The existing quality control system of tobacco company is composed of three parts:
- Part 1: Online inspection of all cigarettes based on Cigarette Inspection System-CIS. As shown in Fig. 1, $\mu_1$ is the specified value of cigarette draw resistance, $X$ needs to meet $X \in [u_1 - n\sigma_1, \mu_1 + n\sigma_1]$, otherwise it is judged as substandard cigarettes and rejected by the equipment online. The draw resistance is essentially a conversion of the pressure value. During the measurement phase, the CIS system clamps the cigarette through a pair of detecting caps, and a constant 1.7kpa of compressed air is ejected. The compressed air passes through the ignition of the cigarette and reaches the cigarette filter, where the detecting cap converts the pressure signal into an electrical signal, which in turn calculates the value of draw resistance.

- Part 2: During the production process, a very small number of cigarettes are randomly selected at regular intervals for manual measurement. Specifically, the quality inspector randomly selects 20 cigarettes from the cigarette-making machine three times a day, and sends them to the testing room for manual measurement.
- Part 3: randomly selected cigarettes from the warehouse are sent to the finished product inspection room for manual measurement. Specifically, the quality inspector randomly selects 50 cigarettes from a production batch (5 randomly selected from 1000 pieces, 1 randomly selected from each piece, 1 randomly selected from each box, 10 cigarettes from each box, 50 cigarettes in total) and sends them to the testing room to manually measure the resistance to smoking, if the deviation is within the permissible range, they are judged to have passed.

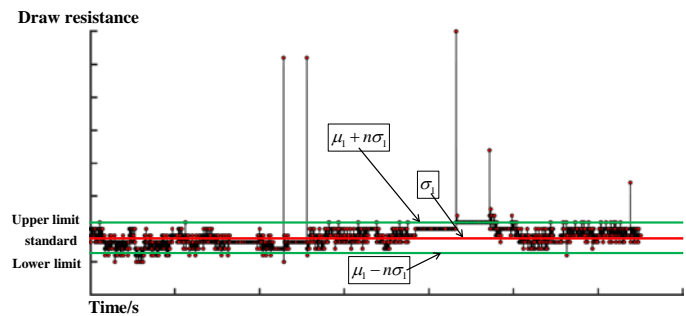

**Figure 1** The basis for the draw resistance to meet the standard

The Part 1's CIS online detection is one of the most important components of cigarette quality monitoring and has an accuracy of 96%, but still suffers from the following problems:

- The CIS is used to measure suction resistance directly and is composed of precision instruments. In the event of mechanical failure, sensor wear, unstable data transmission and other uncertainties, the detection of draw resistance becomes meaningless. Therefore, the worker has to carry out the tedious calibration of the CIS at regular intervals.
- The online monitoring method is single and lacks correlation modelling, furthermore, the monitoring results cannot be used to guide the sampling of Part 2 and 3. As a result, no further fault time periods can be located.

The manual measurement methods for Part 2 and 3 are far less important than Part 1, and suffer from the following problems:

- The cost of manpower. Quality inspectors are required to sample multiple cigarette-making machines at a time, in addition, travel to and from the workshop and testing room daily.
- The cost of time. Manual testing in the testing room is slow, yet the cigarettes are produced very quickly, so the results are not fed back to production in time.
- The cost of hardware. The testing room is required to cover the cost of repair and

replacement of testing instruments.
- Poor reliability of test results. Due to unreliable factors such as the quality inspector's professionalism, skill level and environment, different quality inspectors may test the same cigarette with different results.

In summary, it is generally agreed that random sampling in the second and third sections contributes less to the monitoring of suction resistance[20]. Therefore, there is an urgent need to develop a method that can assist Part 1 and guide Part 2-3, i.e. to achieve correlative modelling of cigarette smoking resistance.

In response to these problems, some researches used regression analysis to monitor draw resistance. Wang[21, 22] and Zhao[23] built a draw resistance prediction model through linear networks with an error of less than 10%, but only linear factors were considered and it takes a long time to measure these factors. To achieve online monitoring, Chang[24] and Chen[25] completed correlation modeling of draw resistance with a prediction accuracy of 96%, which considered a small number of factors that were measured in real-time. However, only linear correlations affecting the draw resistance were considered, such as filler value, moisture content, cigarette weight, circumference, etc. Apart from this, the above did not clean up the raw data and did not allow for true online real-time prediction. To overcome these problems, Pang[26] performed online prediction of cigarette draw resistance based on multi-sensor fusion with a prediction accuracy of 87%, but the large time granularity (1min) resulted in the inability to achieve anomaly localization. Tang[27] used recurrent neural networks to achieve online prediction of draw resistance at a finer temporal granularity (2s), but the lack of expert experience resulted in important factors such as sensor temperature being overlooked and did not guide manual sampling (i.e. Part2-3).

Given the above, this paper proposes a method for cigarette draw resistance monitoring without increasing the hardware cost, as shown in Fig. 2, which is composed of three parts:

- *Step1-In each unit of cigarette rolling, factors that affect the draw resistance are explored.* In the raw material, the VE, the SE, and the MAX units, process indicators as well as control indicators are pre-analyzed, to initially study direct or potential indicators affecting the cigarette draw resistance. Then, the Spearman coefficient and random forest assess their correlations with draw resistance, from which the relevant indicators are further explored.
- *Step2-Draw resistance prediction.* Cigarette draw resistance is predicted based on multi-indicator correlation analysis, to identify substandard cigarettes missed by the CIS.
- *Step3-The validity of cigarette monitoring is verified*. For a given production shift, the distribution characteristics of the number of substandard cigarettes are analyzed, the distribution curves are fitted, and then a probability model of substandard cigarettes being sampled is constructed, which further highlights the usefulness of predictive models.

The chapters of this paper are organized as follows: Chapter 1 discusses the current status and proposes an improved monitoring method for cigarette draw resistance;

Chapter 2, the direct or potential correlation between each indicator and draw resistance is analyzed on three units(VE, SE, and MAX), which the correlation factors of the draw resistance are determined, to construct the draw resistance prediction model; Chapter 3, to further verify the validity of the prediction model of cigarette draw resistance, the probability model of sampling substandard cigarettes is derived; Chapter 4, describes the experiments and results, including data sources, data preprocessing, and experimental parameter settings; Chapter 5, discusses the results and future work.

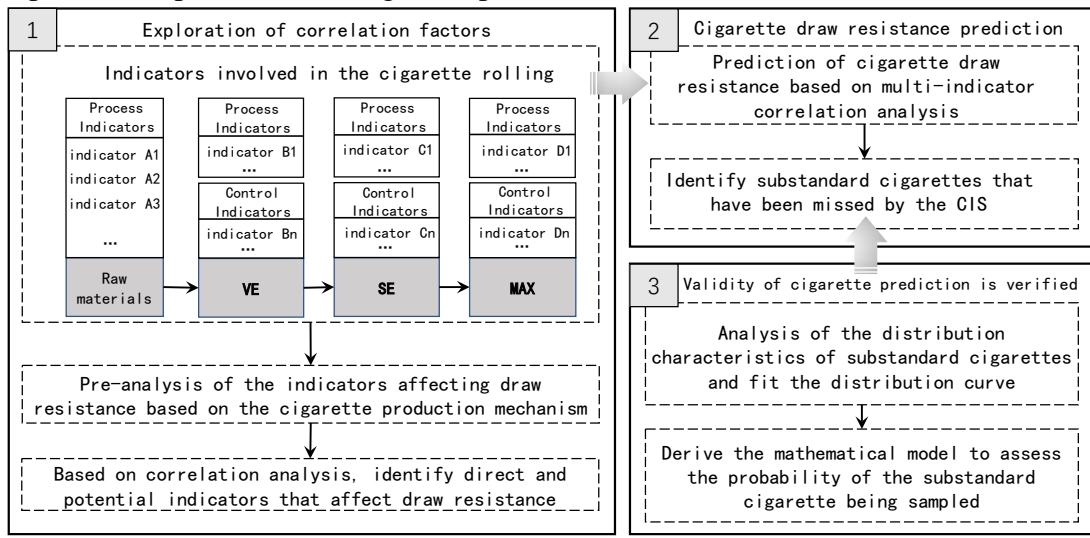

**Figure 2** The prediction-based method for monitoring cigarette draw resistance

2. Prediction-based cigarette draw resistance monitoring.

In this paper, the quality monitoring method for cigarette draw resistance is proposed, as Fig. 2, including the exploration of correlation factors, the cigarette draw resistance prediction, and the validity of cigarette prediction is verified.

2.1 The exploration of correlation factors

2.1.1 Pre-analysis of the indicators affecting draw resistance based on the cigarette rolling production mechanism

The production of cigarette rolling is shown in Fig. 3. The purpose of cigarette rolling is to process the cut tobacco into complete cigarettes. Based on the analysis of the cigarette-forming mechanism in cigarette rolling, the correlation indicators of the raw materials, the cut tobacco supply (VE), the rolling tobacco bundles (SE), and the cigarette-forming (MAX) are explored, to build a prediction model for draw resistance, which enriches the method of draw resistance monitoring or assist the CIS. The flow of the cigarette rolling is shown in Fig. 4.

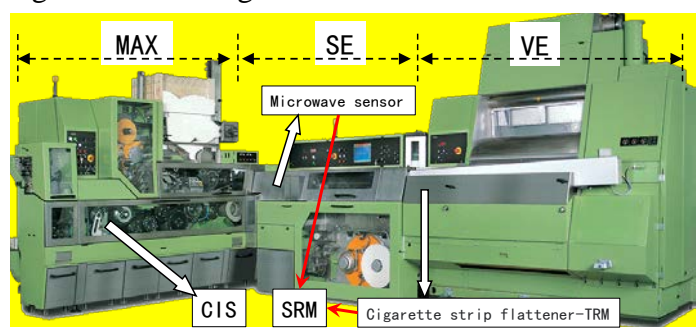

**Figure 3** Cigarette maker (includes the VE, the SE, and the MAX)

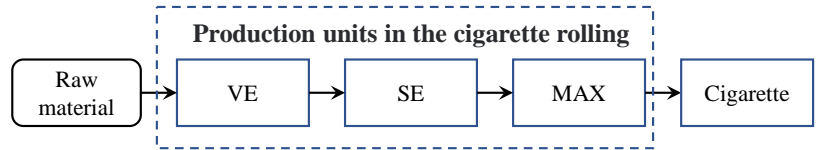

**Figure 4** The flow of the cigarette rolling

2.1.1.1 Factors involved in raw materials

Raw materials are mainly tipping paper and cut tobacco. The tipping paper is mainly concerned with its temperature. The physical indicators of the cut tobacco are: moisture content before using, filling power, long filament rate, froth filament rate, short filament rate, broken filament rate. Cut tobacco needs to be kept at a standard moisture content at all times, so that the best filling power of the tobacco is obtained [28-30]. Filling power is used to measure the degree of filling inside the cigarette, which directly affects the weight of the cigarette, and indirectly affects the draw resistance of the cigarette[31]. The long filament rate, froth filament rate, short filament rate and broken filament rate directly affect the hardness and circumference of the cigarette. If the weight of the cigarette remains the same, the increase in the short filament rate will lead to a decrease in the hardness and diameter of the cigarette[32]. The different structures of the tobacco filaments are shown in Fig. 5.

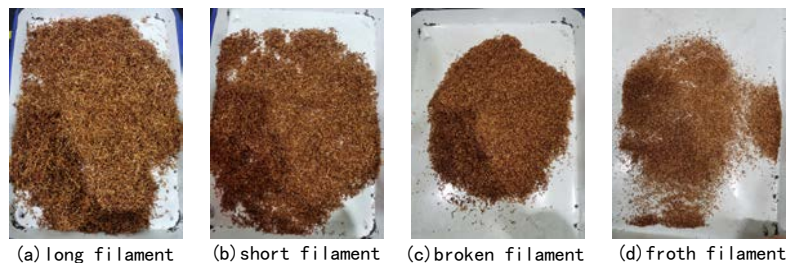

**Figure 5** Different structures of the tobacco filaments

2.1.1.2 Factors involved in the VE

In the VE processing unit, the feeder makes the cut tobacco into bundles and delivers them to the SE, Fig. 6 shows the cut tobacco feeder in the VE.

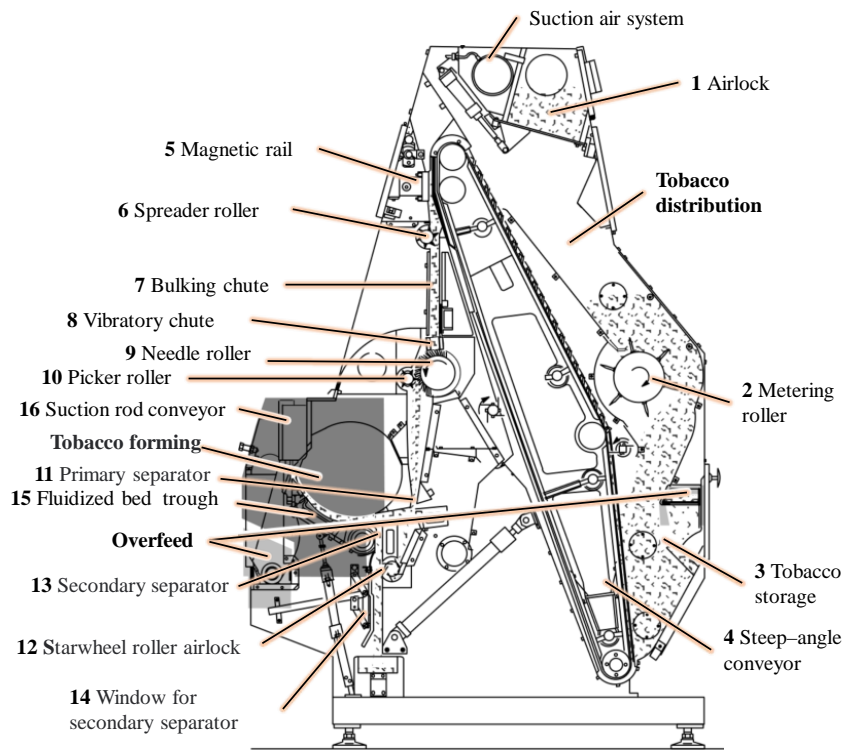

**Figure 6** Cut tobacco feeder in the VE

In Fig. 6. (1) sends the cut tobacco to the VE process, and after (1)-(11), and the tobacco is sent to (16) by the airflow of (11) and (13). Eventually, the tobacco on (16) is sent into (17) by the airflow and is adsorbed on the suction belt to the smoking gun barrel of the SE. Before the tobacco reaches the smoking gun barrel, it will go through the trimmer disks(Fig. 7-E) to cleave off the excess tobacco from the tobacco bundle, to ensure the weight of the tobacco and indirectly control the draw resistance, which is a system feedback control.

Tobacco transmission relies on an air circulation system with the steam distribution box as the core unit, and the air consumption units are (11), (13) and (16), then (11) and (13)'s purposes are remove tobacco stems. The (15)'s upper pressure directly determines the amount of adsorbed tobacco. The tobacco is adsorbed by suction rod conveyor vacuum pressure on the suction tape, and the tape will be torn when the tension pressure is too high. The tobacco temperature is measured by the sensor in (7), and the tobacco temperature directly affects the calculation of the tobacco density in the subsequent SE unit, which in turn affects the calculation of the tobacco weight and the draw resistance.

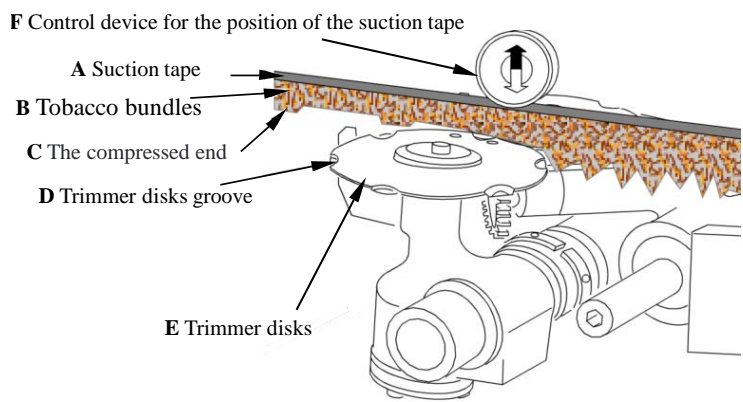

**Figure 7** Trimmer(TRM)

Trimmer disks, suction tape are the main devices of the cigarette trimming system (TRM), as shown in Fig. 7. the weight of the cigarette is determined by the position of (A), which is controlled by (F), as follows:

| Suction tape position | Amount of tobacco | Cigarette weight | Draw resistance |
|---|---|---|---|
| Up | Increase | Increase | Increase |
| Down | Decrease | Decrease | Decrease |

The purpose of designing D on E is to ensure a higher density of tobacco at the end of the strip, which is used to compensate for loose and shaken tobacco during transportation.

In summary, the factors that directly and indirectly affect the draw resistance in the VE process: pressure of air distributor boxes, cut tobacco temperature in tobacco storage, the pressure of primary separator, the upper pressure of fluidized bed trough, the suction rod conveyor vacuum pressure, the suction tape tension pressure, the suction tape position, the target weight, the cigarette weight, and the position of cigarette compaction end.

2.1.1.3 Factors involved in the SE

The SE processes the bundles of tobacco delivered by the VE unit into sticks. The tobacco is wrapped in paper and cut into strips, and the weight of the tobacco is measured before being sent to the MAX filter loading machine. The tobacco is wrapped in paper and cut into strips, and the weight of the strips is measured before being sent to the MAX.

The core part of the SE is the SRM, which measures the cigarette weight through a microwave (MW) measuring head (MIDAS) (Fig. 8). In daily production, the cigarette draw resistance is changed by adjusting the cigarette weight, so the indicators related to the cigarette weight are also potential factors that affect the draw resistance[31-33]. The cigarette circumference is related to the position of the pressure plate.

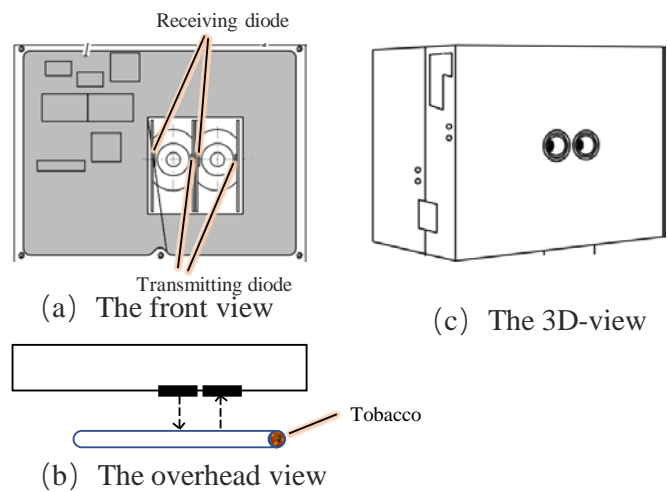

(a) The front view  (c) The 3D-view

(b) The overhead view

**Figure 8** The MW-MIDAS

MW-MIDAS provides all measured values for the SRM. The voltage of the microwave resonator in the MW-MIDAS varies with the tobacco density. MW-MIDAS electrical characteristics are susceptible to temperature, so the temperature of the surrounding hardware, such as microwave components, must be considered[34].

In addition, if the soldering iron temperature is too low, the gap at the cigarette paper adhesion will become larger, which will directly affect the cigarette airtightness. The SE belt drives the cigarette paper and tobacco bundle through the smoking gun barrel, so the belt must be tensioned as tightly as possible or the paper will wobble.

In summary, the factors that directly and indirectly affect the draw resistance in the SE unit: cigarette circumference, position of the pressure plate, the number of overweight cigarettes, number of light cigarettes, number of soft cigarettes, number of hard cigarettes, the solder iron temperature, the power supply temperature, circuit board temperature, the shell temperature, the resonator temperature, the diode temperature, and the SE belt tension, etc. It is worth noting that, based on expert experience, the temperature of the electronic components is one of the potential factors affecting the measurement accuracy, which is one of the differences between this paper and the Pang[26].

2.1.1.4 Factors involved in the MAX

The filter tipping machine of the MAX(Fig. 9), splices the filter with the cigarette sent by SE, which forms the whole cigarette. In addition, some indicators be measured by the CIS, such as air leakage, ventilation strength, and draw resistance of every cigarette.

As shown in Fig. 9, the "Supply cigarettes without filter" receives cigarette from the SE, and the "Supply filter rod" provides filter rod, in addition, the "Supply tipping paper" provides tipping paper, and then, both cigarette and filter rod are wrapped by tipping paper in the "cigarette production system with filters", which to form the whole cigarette. The subsequent CIS monitors the air tightness and density of the cigarette online. In monitoring the airtightness of a cigarette, there is a constant pressure airflow from the end of the cigarette to the pressure sensor at the filter. The compressed air pressure in the MAX is controlled by the "Control of air intake pressure".

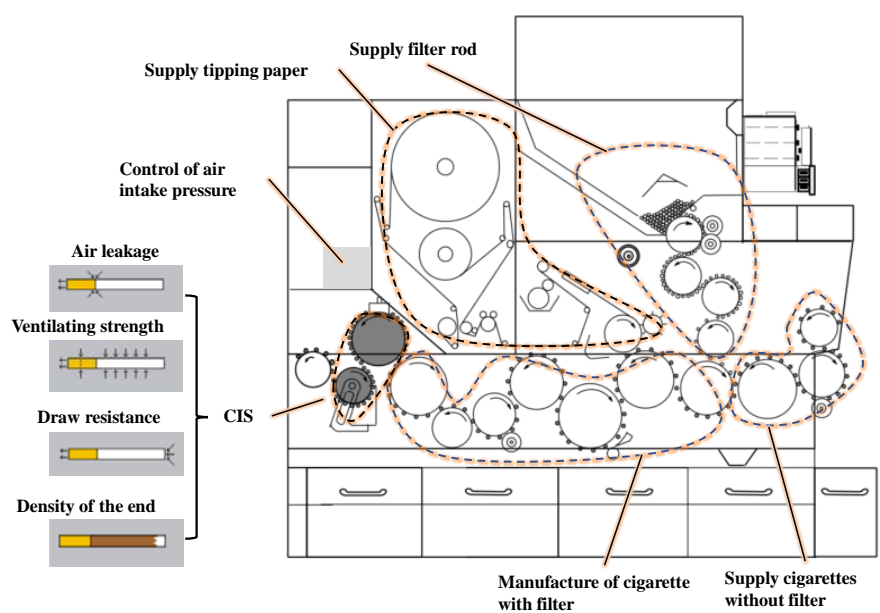

**Figure 9** The filter tipping machine of the MAX

The reasons for poor cigarette airtightness:
1. Changes in the temperature of the rolling plate and the tipping paper.
2. Changes in sensor performance due to changes in compressed air.
3. Change in cigarette weight.

In summary, the factors that directly and indirectly affect the draw resistance in the MAX unit: rolling plate temperature, intake air pressure, air source pressure, cigarette end density, cigarette weight, air leakage, ventilating strength, etc.

2.1.1.5 Results of pre-analysis of affecting indicators

Combining the pre-analysis results of the VE, SE, and MAX units, the affecting indicators of cigarette draw resistance based on cigarette production mechanism is mined, and the results are shown in Fig. 10. a total of 82 indicators are retained for pre-analysis, including linear and non-linear correlations. Since draw resistance is closely related to cigarette weight[24], in this paper, cigarette weight-related indicators will also be fully considered to predict draw resistance. Indicators that are indirectly related to the draw resistance, such as those related to air pressure and hardware temperature, are considered. However, the air source pressure and intake pressure in MAX are almost constant and are not used as draw resistance prediction.

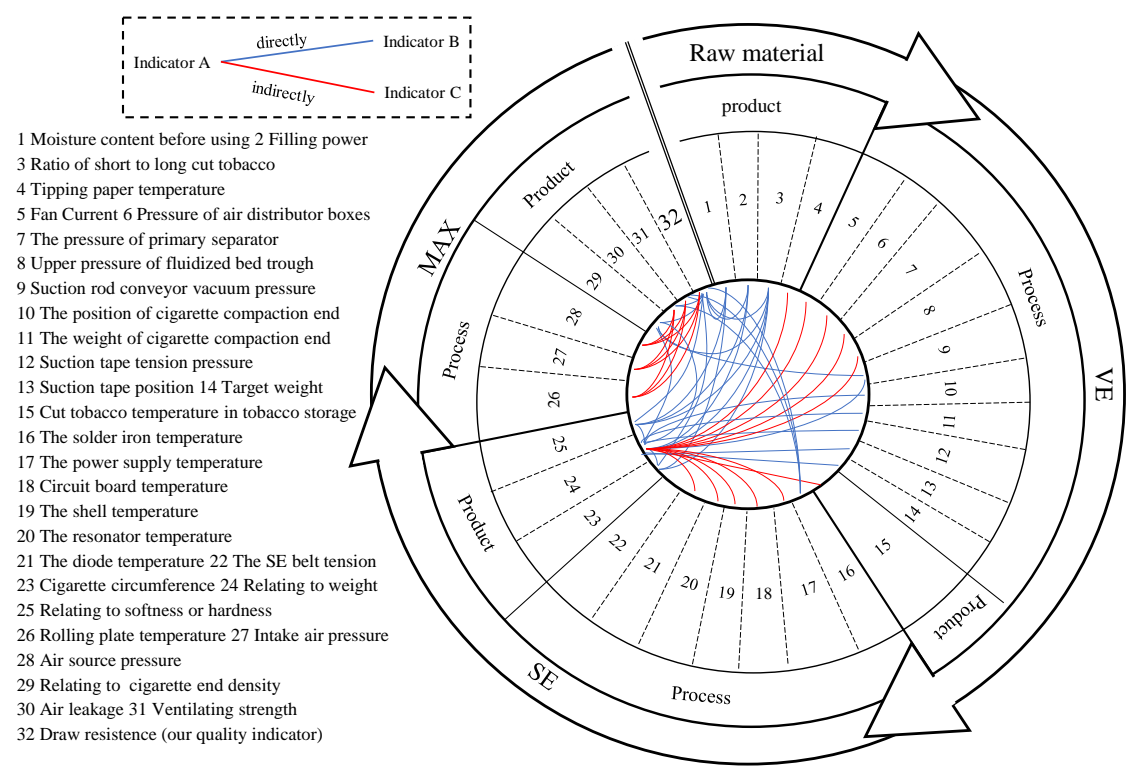

1 Moisture content before using 2 Filling power
3 Ratio of short to long cut tobacco
4 Tipping paper temperature
5 Fan Current 6 Pressure of air distributor boxes
7 The pressure of primary separator
8 Upper pressure of fluidized bed trough
9 Suction rod conveyor vacuum pressure
10 The position of cigarette compaction end
11 The weight of cigarette compaction end
12 Suction tape tension pressure
13 Suction tape position 14 Target weight
15 Cut tobacco temperature in tobacco storage
16 The solder iron temperature
17 The power supply temperature
18 Circuit board temperature
19 The shell temperature
20 The resonator temperature
21 The diode temperature 22 The SE belt tension
23 Cigarette circumference 24 Relating to weight
25 Relating to softness or hardness
26 Rolling plate temperature 27 Intake air pressure
28 Air source pressure
29 Relating to cigarette end density
30 Air leakage 31 Ventilating strength
32 Draw resistence (our quality indicator)

**Figure 10** The results of pre-analysis of affecting indicators

2.1.2 Mining affecting factors based on correlation analysis

The scale of the indicators involved in cigarette reel production is large, as shown in Figure 10, and their being used directly in the prediction model would result in the optimal features not being accurately mined. Correlation analysis allows feature filtering and avoids unimportant features in high-dimensional data. Current cigarette draw resistance correlation analysis only uses simple linear correlations such as Pearson's coefficient to assess the degree of correlation and lacks consideration of nonlinear potential indicators. Currently for cigarette draw resistance correlation analysis, only simple linear correlations such as Pearson coefficient are used to assess the correlation, lacking consideration of nonlinear potential indicators. Based on the pre-analysis (Section 2.1.1), the Spearman coefficient as well as the random forest are used and the direct (linear correlation), and potential (nonlinear correlation) correlations affecting the draw resistance are calculated.

The linear correlation of the indicators was assessed based on the Spearman coefficient. The importance of the indicators was taken as absolute values, and the results are shown in Figure 11. For the parameters of the RF, the number of trees is 20, and the number of leaves is 5, as in Figure12. The nonlinear correlations of the indicators are evaluated based on random forest (RF), and the results are shown in Figure 13.

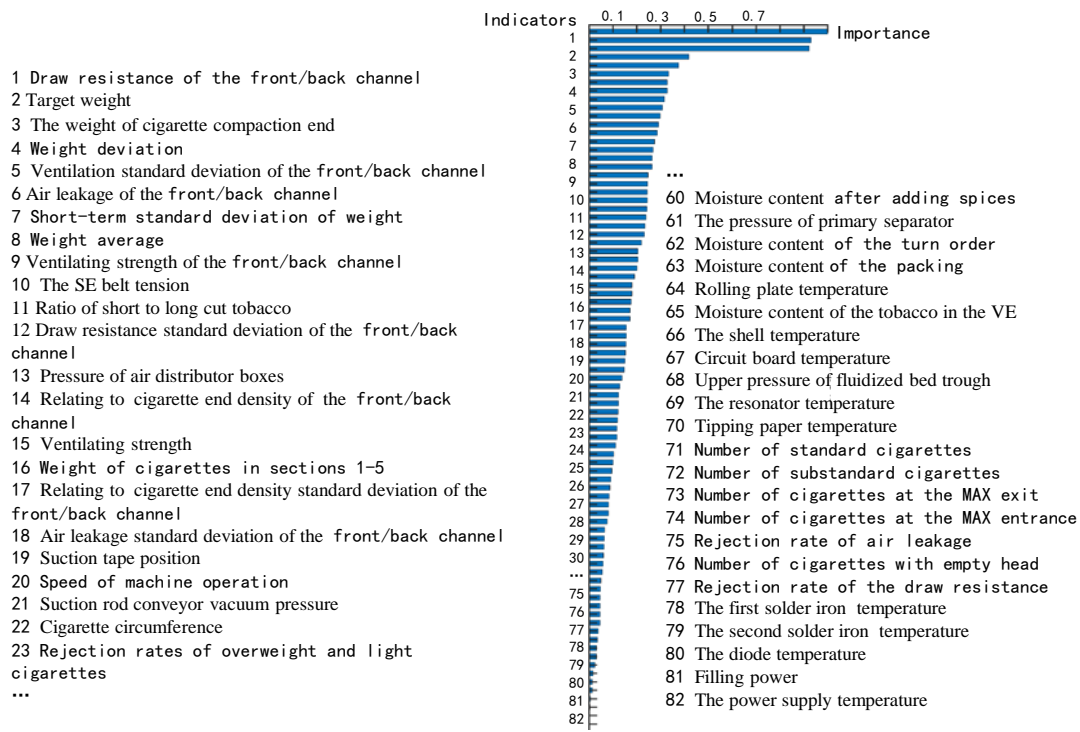

**Figure 11** Direct influence factors (linear correlation) based on the Spearman coefficient

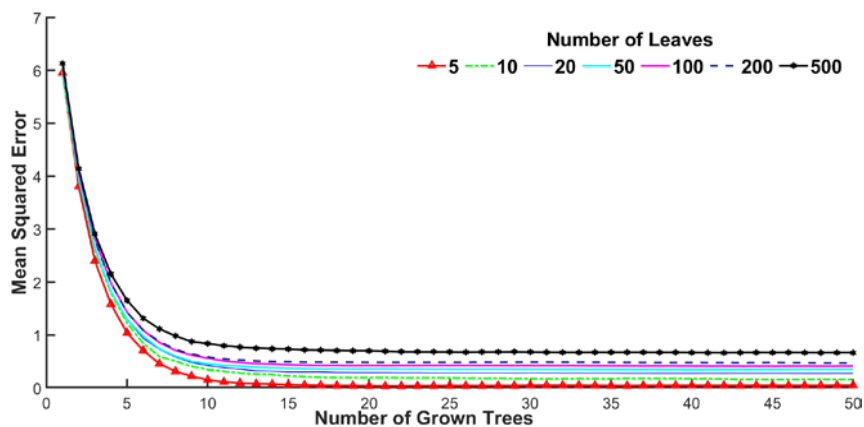

**Figure 12** Determine the optimal parameters for RF($Trees = 20, leaves = 5$)

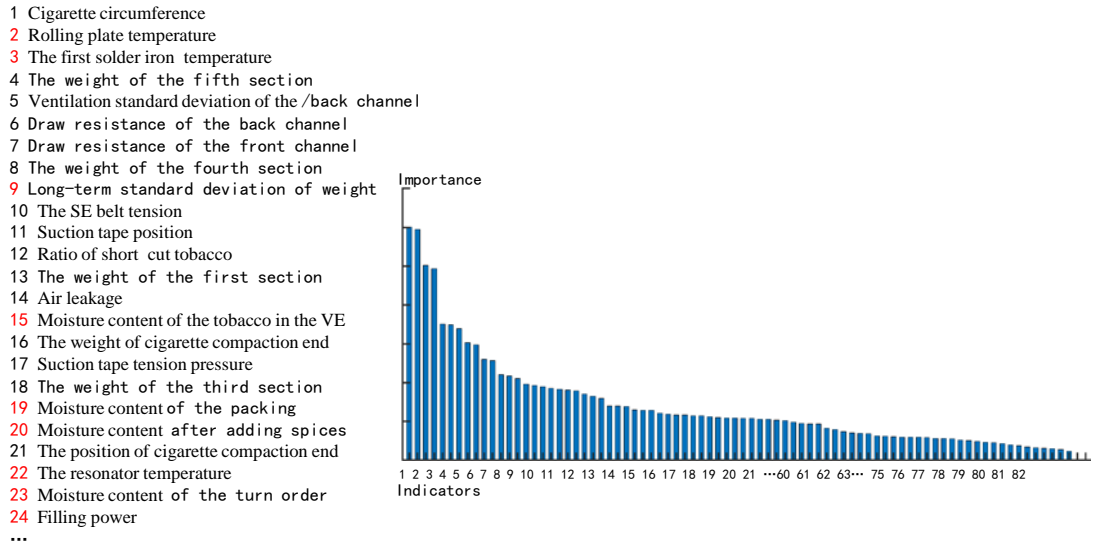

1 Cigarette circumference
2 Rolling plate temperature
3 The first solder iron temperature
4 The weight of the fifth section
5 Ventilation standard deviation of the /back channel
6 Draw resistance of the back channel
7 Draw resistance of the front channel
8 The weight of the fourth section
9 Long-term standard deviation of weight
10 The SE belt tension
11 Suction tape position
12 Ratio of short cut tobacco
13 The weight of the first section
14 Air leakage
15 Moisture content of the tobacco in the VE
16 The weight of cigarette compaction end
17 Suction tape tension pressure
18 The weight of the third section
19 Moisture content of the packing
20 Moisture content after adding spices
21 The position of cigarette compaction end
22 The resonator temperature
23 Moisture content of the turn order
24 Filling power
...

**Figure 13** Indirect influence factors (non-linear correlation) based on the RF

The key to factor selection is to identify those factors that affect the draw resistance. Due to constraints in computational power or model fitting ability, traditional methods only rely on expert experience to select those factors with strong linear correlation, such as [21], [23], [24], [25], and only the Ventilating strength, the Filler value, the Moisture content, and the Cigarette weight are considered. Moreover, fewer samples are used for experiments, resulting in the inability to achieve online prediction, difficulty in improving prediction accuracy, and poor model generalization ability. The quality monitoring of the cigarette rolling process involves about 200 factors and 2T of data per day, which needs to be fully used for correlation modeling to mine the data value. With the development of deep learning and the improvement of computational power, models with better fitting ability are used for siphon resistance prediction, such as [26], [27], but there is no systematic analysis of the production process, resulting in the Temperature being ignored, which ultimately limits the accuracy of draw resistance prediction. As in Figure 11, the top 30% (25 factors) of indicators for linear correlation are considered as direct influences in this paper. In Figure 13, the importance of some potential indicators increased significantly in the non-linear correlation analysis, and the top 24% (14 factors) of indicators in the non-linear correlation (after removing the direct influencing factors) are used in this paper. The final indicators used as predictors of cigarette smoking resistance are shown in Table 5.

2.2 Draw resistance prediction model based on multi-indicator

Based on the analysis of the correlation factors of cigarette draw resistance, the draw resistance prediction model is constructed (Fig. 13).

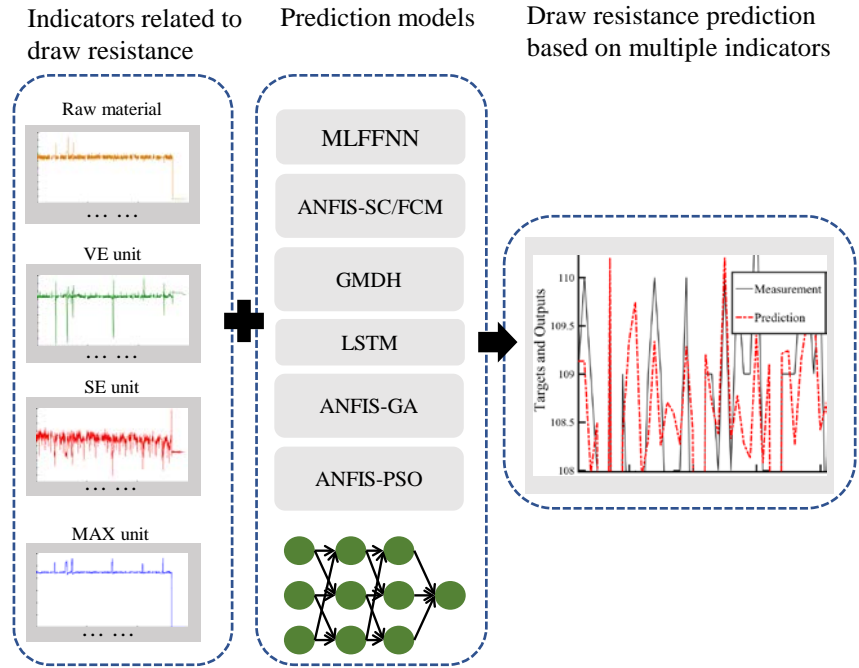

**Figure 13** Draw resistance prediction based on multiple indicators

2.2.1 Multi-layer feedforward neural networks-MLFFNN

Multi-layer feedforward neural networks are the most commonly used artificial neural networks, including perceptron networks, BP networks, and RBF networks. The BP network used in this paper is a typical application of the back propagation learning algorithm in MLFFNN[35], and a variety of training functions are used for comparative analysis, including gradient descent training function (traingd), Bayesian regularization training function (traingbr), Levenberg-Marquardt algorithm (trainlm), Plwell-Beale algorithm (traincgb) and conjugate gradient algorithm (trainscg).

2.2.2 Adaptive neuro fuzzy inference system-ANFIS

Adaptive neuro fuzzy system-ANFIS, which integrates the learning algorithm of artificial neural network (ANN) and the concise form of fuzzy inference (FIS), the training process of the model can be reduced to the process of adjusting parameters by error back propagation with least squares. Two inference methods exist for ANFIS, Mamdani and Sugeno, and Sugeno used in this paper is more suitable for optimization and adaptive. ANFIS contains three main parts: fuzzification, fuzzy inference, and defuzzification[36]. Fig. 14 shows the simple ANFIS model having two inputs and one output, two fuzzy rules are used, as in Eqs. (1) and (2).

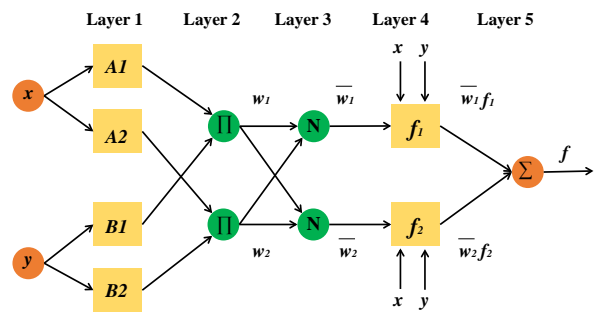

**Figure 14** The simple structure of ANFIS model

Rule1: if $x$ is A1 and $y$ is B1, then $f = p_1 x + q_1 y + r_1$　　　　(1)

Rule2: if $x$ is A2 and $y$ is B2, then $f = p_2 x + q_2 y + r_2$　　　　(2)

As shown in Fig. 14, the structure of ANFIS has five layers, and the first and second square nodes require parameter learning, which is the core part of the model, and the functions of each layer are described as follows:

Layer 1(Fuzzification layer): The layer is the fuzzification layer, $A_i$、$B_i$ are the fuzzy sets, which fuzzes the input variables and outputs the membership degree of the fuzzy set to the layer 2. And the output($O_i^1$) of the layer can show the extent to which $x$ and $y$ belong to $A_i$ and $B_i$, as Eqs. (3) and (4). $\mu_{A_i}$ and $\mu_{B_i}$ are the Gaussian functions, as Eqs. (5) and (6).

$$O_i^1 = \mu_{A_i}(x) \quad i = 1, 2 \tag{3}$$

$$O_i^1 = \mu_{B_i}(y) \quad i = 1, 2 \tag{4}$$

$$\mu_{A_i} = \exp\left\{-\left[\left(\frac{x - c_i}{a_i}\right)^2\right]^{b_i}\right\} \tag{5}$$

$$\mu_{B_i} = \exp\left\{-\left[\left(\frac{y - c_i}{a_i}\right)^2\right]^{b_i}\right\} \tag{6}$$

where $x$ and $y$ are input variables. $\{a_i、b_i、c_i\}$ are premise parameters, which need be trained.

Layer 2 (Rule layer): The nodes in this layer are responsible for the multiplication of the input signals, and the membership degree of each node input represents the release strength of a fuzzy rule.

$$O_2^i = w_i = \mu_{A_i}(x) \bullet \mu_{B_i}(y) \quad i = 1, 2 \tag{7}$$

Layer 3 (Normalization layer): This layer normalizes the output of the layer 2, and describes the proportion of the rule's firing (expressed as probability) in the overall rule base.

$$O_3^i = \overline{w}_i = \frac{w_i}{w_1 + w_2} \quad i = 1, 2 \tag{8}$$

Layer 4 (Defuzzification layer) is used to calculate the each rule's output.

$$O_4^i = \overline{w}_i f = \overline{w}_i (p_i x + q_i y + r_i) \quad i = 1, 2 \tag{10}$$

where the $\{p_i、q_i、r_i\}$ are the consequent parameters, which be trained.

Layer 5 (sum layer): Calculating the total output of all input signals.

$$O_5^i = \sum \overline{w}_i f_i = \sum w_i f_i / \sum w_i \quad i = 1, 2 \tag{11}$$

ANFIS model can be improved by particle swarm optimization (PSO)[37], and genetic algorithm (GA)[38]. The process of training the ANFIS model is to reduce the network error by seeking the optimal premise parameters (Eqs. (5)-(6)), and the consequent parameters (see Eq. (10)). PSO and GAs can help to find the global optimum of the above parameters[39].

2.2.3 Long short term memory network-LSTM

LSTM belongs to a type of recurrent neural network that is used in time-series prediction, and LSTM is capable of multiple inputs and multiple outputs. LSTMs perform well in applications with strong spatio-temporal properties, such as traffic flow control[40], climate prediction[41, 42], motion pose recognition [43], and video processing [44].

2.2.4 Group method of data handling-GMDH

GMDH automatically finds the correlation of each metric, and the neurons within the same layer are combined two by two and intermediate models are generated (internal criterion). Then, based on the external criterion, the intermediate models are filtered and the models that satisfy the conditions become the next layer of neurons, and the above process is repeated until the optimal complexity of the network is obtained[45]. Fig. 15 shows the simple structure of GMDH.

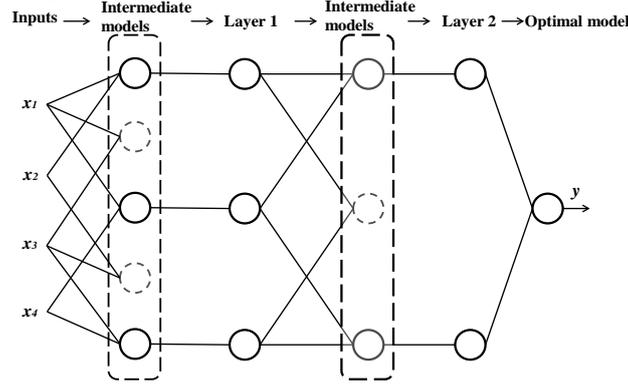

**Figure 15** The simple structure of GMDH model

3. Building a probability model

To further highlight the practicality of the prediction model, this paper innovatively builds a probability model. Based on the prediction results, the distribution characteristics of substandard cigarettes are analyzed, the period at which substandard cigarettes are concentrated is located, and the probability model of substandard cigarettes being sampled is derived.

If the probability of substandard cigarettes being sampled is significantly increased, the usefulness of the prediction model is further verified. The practical value of the prediction model is further validated if the probability of substandard cigarettes being sampled is significantly increased (It is worth noting that the method is highly generalizable. It is strongly enlightening and instructive for other areas of cigarette quality monitoring, such as traceability and focused sampling).

There are many methods that can be used to assess the actual value of a predictive model, and this paper uses the probability $P$ of a substandard cigarette being sampled as an example:

$$P = 1 - \frac{C_m^0 * C_{Y-m}^Z}{C_Y^Z} = \frac{A_{Y-m}^Z}{A_Y^Z} \quad (12)$$

where $P$ is the probability of a substandard cigarette being sampled. $Y$ is the total number of cigarettes involved in the sampling. $Z$ is the number of cigarettes sampled. If $Y$ satisfies $Y \gg m, Z$:

$$P \approx 1 - \left(\frac{Y-m}{Y}\right)^Z \quad (13)$$

The existing sampling method lacks a priori knowledge about the characteristics of the distribution, so random sampling is used, and the default distribution is random and uniform:

$Y$ : the total number of cigarettes produced in a work shift, and is set to $Y_1$ ;

$$Y_1 = \sum_{id=1}^{j_1} y_{id} \quad (14)$$

where $id$ is the serial number of a period in the shift. $y_{id}$ is the number of cigarettes produced in the time period with the serial number $id$. $j_1$ is the number of periods included in the shift. Each period has the same length, and the number of cigarettes produced in a period is $y_s$.

$$Y_1 = j_1 * y_s \quad (15)$$

$Z$: determined by the total number of cigarettes involved and is a fixed constant $Z_1$;
$m$: number of all substandard cigarettes during the shift is set to $m_1$.

In summary, the probability $P_{old}$ of a substandard cigarette being sampled is as follows:

$$P_{old} = 1 - \frac{C_{m_1}^0 * C_{Y_1-m_1}^Z}{C_{Y_1}^{Z_1}} = 1 - \frac{A_{Y_1-m_1}^{Z_1}}{A_{Y_1}^{Z_1}} \approx 1 - \left(\frac{Y_1 - m_1}{Y_1}\right)^{Z_1} = 1 - \left(\frac{j_1 * y_s - m_1}{j_1 * y_s}\right)^{Z_1} \quad (16)$$

Based on the prediction results of cigarette draw resistance, the distribution characteristics of substandard cigarettes can be directly obtained. The potential distribution characteristics include uniform distribution, normal distribution, and t-distribution, etc. In this paper, only the normal distribution is considered. As shown in Figure 16, the horizontal coordinate is the serial number of each period (no strict time order) and the vertical coordinate is the number of substandard cigarettes. The number of substandard cigarettes is fitted to a normal curve, and the serial numbers of all periods satisfy $ID \sim N(\mu, \sigma^2)$. The standard deviation is $\sigma$ and the mean $\mu$ is the serial number of the period with the most substandard cigarettes.

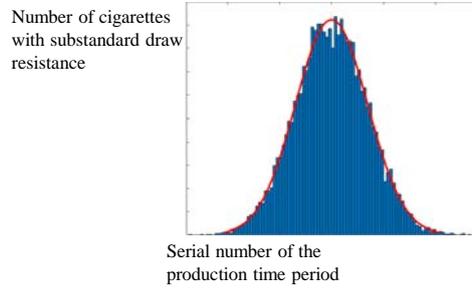

**Figure 16** Distribution characteristics of substandard cigarettes

In summary, the $Y$ in Eqs. (12) and (13) is no longer considered as all the cigarettes in the shift, but the cigarettes in the set of periods that satisfy $ID \in (\mu - n\sigma, \mu + n\sigma)$. Therefore, the Eqs. (12) and (13) is optimized as follows:

$Y$: The total number of cigarettes produced in periods satisfying $ID \in (\mu - n\sigma, \mu + n\sigma)$, and is set to $Y_2$;

$$Y_2 = \sum_{id=1}^{ID} y_{id} \quad (17)$$

where $id$ is a subset of $ID$. $ID$ is the set of serial numbers satisfying $ID \in (\mu - n\sigma, \mu + n\sigma)$, the number of periods included is $j_2$.

$$j_2 = 2n\sigma, \exists n \in N^* \Rightarrow j_2 < j_1 \quad (18)$$

$Z$: determined by the total number of cigarettes involved and is a fixed constant $Z_2$, $Z_2 = Z_1$.

$m$ : the number of substandard cigarettes produced in the period $X$ satisfying $X \in (\mu - n\sigma, \mu + n\sigma)$.

$$m_2 = \Phi(n) * m_1 \quad (19)$$

$$\Phi(n) = P(\mu - n\sigma < \mu < \mu + n\sigma) \quad (20)$$

$$P(a < x \leq b) = \int_a^b \frac{1}{\sqrt{2\pi} * \sigma} e^{-\frac{(x-\mu)^2}{2\sigma^2}} dx \quad (21)$$

In summary, the optimized probability $P_{new}$ of a substandard cigarette being sampled in this paper is as follows:

$$P_{new} = 1 - \frac{C_{m_2}^0 * C_{Y_2 - m_2}^Z}{C_{Y_2}^{Z_2}} = 1 - \frac{A_{Y_2 - m_2}^{Z_2}}{A_{Y_2}^{Z_2}} \approx 1 - \left(\frac{Y_2 - m_2}{Y_2}\right)^{Z_2} = 1 - \left(\frac{j_2 * y_s - m_2}{j_2 * y_s}\right)^{Z_2} \quad (22)$$

The probability of substandard cigarettes being sampled is raised by $\Delta P$:

$$\Delta P = P_{new} - P_{old} > 0 \quad (23)$$

Based on Eqs. (16), (19), (20), (21) and (22), Eq (23) is as follows:

$$\Delta P \approx \left(\frac{j_2 * y_s - m_2}{j_2 * y_s}\right)^{Z_2} - \left(\frac{j_1 * y_s - m_1}{j_1 * y_s}\right)^{Z_1} = \left(\frac{2n\sigma * y_s - \Phi(x)m_1}{2n\sigma * y_s}\right)^{Z_1} - \left(\frac{j_1 * y_s - m_1}{j_1 * y_s}\right)^{Z_1} \quad (24)$$

4. Experiment and discussion

Firstly, the source of the original data is introduced, secondly, the data is pre-processed, and finally, the training of the prediction model is implemented and the accuracy is verified.

4.1 Data source

In the VE, SE, and MAX processing units, the number of collected indicators is greater than 100, which includes process indicators, product indicators, etc. The storage frequency of each indicator is 2s, to reduce the storage cost during the actual storage, the value of the indicator is stored only when it changes. Therefore, the storage moments are different for different indicators, as shown in Table 1.

**Table 1** Original data in database

| Indicator A | | Indicator B | |
|---|---|---|---|
| Time Stamp | Value | Time Stamp | Value |
| 2022/02/07 10:51:30.000 | 7002 | 2022/02/07 10:51:30.000 | 58 |
| 2022/02/07 10:51:32.000 | 8209 | 2022/02/07 10:52:06.000 | 59 |
| 2022/02/07 10:51:36.000 | 9490 | 2022/02/07 10:52:16.000 | 57 |
| 2022/02/07 10:51:44.000 | 9492 | 2022/02/07 10:52:18.000 | 60 |
| 2022/02/07 10:51:46.000 | 9498 | 2022/02/07 10:53:44.000 | 63 |
| 2022/02/07 10:51:54.000 | 9496 | 2022/02/07 10:53:46.000 | 62 |
| … | … | … | … |
| 2022/04/29 14:14:02.000 | 9496 | 2022/04/29 15:28:32.000 | 63 |
| 2022/04/29 15:30:34.000 | 9798 | 2022/04/29 15:30:34.000 | 61 |

More than one hundred indicators were collected from February 7 to April 29, 2022, and the collection points were distributed in the VE, SE, and MAX production units. This paper focuses on the indicators related to cigarette draw resistance (see Section 2.1).

4.2 Data pre-processing

The data size of each indicator is not the same due to the storage rules in the database of tobacco companies (see in Table 1), therefore, the data is first filled, to ensure that different indicators have the same size of data and are collected at the same point in time. After filling, each indicator contains 3554194 values, and the whole sample size is [3554194 × number of indicators], the first value is recorded at "2022/02/07 10:51:30", and the last value is recorded at "2022/ 04/29 15:30:34". The filling work is shown in Table 2.

**Table 2** Data filling work

| Indicator A- Original data | | Indicator A- Filling data | |
|---|---|---|---|
| Time Stamp | Value | Time Stamp | Value |
| 2022/02/07 10:51:30.000 | 7002 | 2022/02/07 10:51:30.000 | 7002 |
| 2022/02/07 10:51:32.000 | 8209 | 2022/02/07 10:51:32.000 | 8209 |
| 2022/02/07 10:51:36.000 | 9492 | 2022/02/07 10:51:34.000 | 8209 |
| 2022/02/07 10:51:44.000 | 9494 | 2022/02/07 10:51:36.000 | 9492 |
| 2022/02/07 10:51:46.000 | 9498 | 2022/02/07 10:51:38.000 | 9492 |
| 2022/02/07 10:51:54.000 | 9498 | 2022/02/07 10:51:40.000 | 9492 |
| 2022/02/07 10:51:56.000 | 9497 | 2022/02/07 10:51:42.000 | 9492 |
| 2022/02/07 10:52:04.000 | 9497 | 2022/02/07 10:51:44.000 | 9494 |
| … | … | … | … |

In the data pre-processing, we find that the original data have a serious problem: the original data might be wrong.

For example, the indicator "Cumulative hours worked per shift", which records the cumulative hours worked by workers in a shift. Two shifts are scheduled for production each day, the morning shift from "7:00am-3:30pm" and the mid-shift from "3:30pm-11:30pm", and the value is not zero only during working hours. However, the original data shows that people start work at "03:25:46", which is clearly an incorrect data that does not match the reality, as shown in Table 3.

In addition, similar errors are found in other indicators during the data pre-processing phase (e.g. the indicator "Total cigarettes produced per shift", where theoretically the rate of cigarette production is determined by the machine performance and is relatively stable, but the recorded original data shows a sudden increase in the number of cigarettes).

There are many other similar data quality issues, which are not listed in this paper. Therefore, the indicator "resistance draw resistance" may also have problems such as low data quality. The cause of low data quality may be the instability of hardware such as communication equipment or the CIS, so it is significant to predict the value of draw resistance based on multi-indicator, to enrich the methods of draw resistance monitoring.

**Table 3** Original data-indicator "Cumulative hours worked per shift"

| Cumulative hours worked per shift | |
|---|---|
| Time Stamp | Value/s |
| 2022/02/10 23:35:50.000 | 23448 |
| 2022/02/11 03:25:40.000 | 23451 |
| 2022/02/11 03:25:42.000 | 23453 |
| 2022/02/11 03:25:46.000 | 6 |
| 2022/02/11 03:25:48.000 | 8 |
| 2022/02/11 03:25:50.000 | 10 |
| 2022/02/11 03:25:52.000 | 12 |
| 2022/02/11 03:25:54.000 | 14 |

The final step of preprocessing is to obtain experimental samples from the filling data. Except for the weekdays from 7:30 am-11:30 pm, there is no production during the rest of the day, therefore, the period of production needs to be intercepted from the filling data. The indicator " Total cigarettes produced per shift " records the total number of cigarettes produced in a shift (about 10,000 cigarettes produced by a machine in 60 seconds). The total number of cigarettes suddenly changes from tens of thousands to zero, and the state indicates that the person is off work. And the state of increasing from zero to within 2,000 in two seconds, which indicates the person is at work.

By setting constraints (e.g., error data for cigarette jumps of tens of thousands is set to 0) and other processing, the filling data is eventually divided into 100 shifts (see Table 4). The commuting time period in Table 4 is used to intercept the data of the remaining indicators as experimental data. It is worth mentioning that workers have the habit of starting or finishing work early, and the working time divided by our method is the actual work time.

**Table 4** Accurate commute times

| Go to work | Get off work |
|---|---|
| '2022-02-08 06:41:16' | '2022-02-08 15:26:10' |
| '2022-02-08 15:26:50' | '2022-02-08 23:36:10' |
| '2022-02-09 06:44:36' | '2022-02-09 15:26:10' |
| '2022-02-09 15:26:16' | '2022-02-09 23:30:10' |
| ...... | ....... |
| '2022-04-25 06:45:34' | '2022-04-25 15:27:50' |
| '2022-04-25 15:27:58' | '2022-04-25 23:40:44' |
| '2022-04-26 06:45:50' | '2022-04-26 08:47:28' |
| '2022-04-27 06:46:12' | '2022-04-27 15:27:48' |
| '2022-04-27 15:30:00' | '2022-04-27 23:37:46' |
| '2022-04-28 06:44:24' | '2022-04-28 14:06:10' |

4.3 Experimental data

Taking the production data of one shift as an example, a sliding window is used, and the window size is set to 1, which to include as much information as possible about the original features, and the single sample consisted of the relevant indicators and cigarette smoking resistance. The relevant indicators include $n_1$ raw material indicators, $n_2$ VE indicators, $n_3$ SE indicators, $n_4$ MAX indicators, 1 cigarette draw resistance value at the moment. The training samples are in the form of $[A1,...,An_1, B1,...,Bn_2, C1,...,Cn_3, D1,...Dn_4, Y]$, and details of the sample structure are shown in Figure 14.

The sample of this experiment covers 100 shifts over 3 months, of which 85% of the data is used for training and validation, and 15% is the test set. The indicators used in the experiment are shown in Table 5.

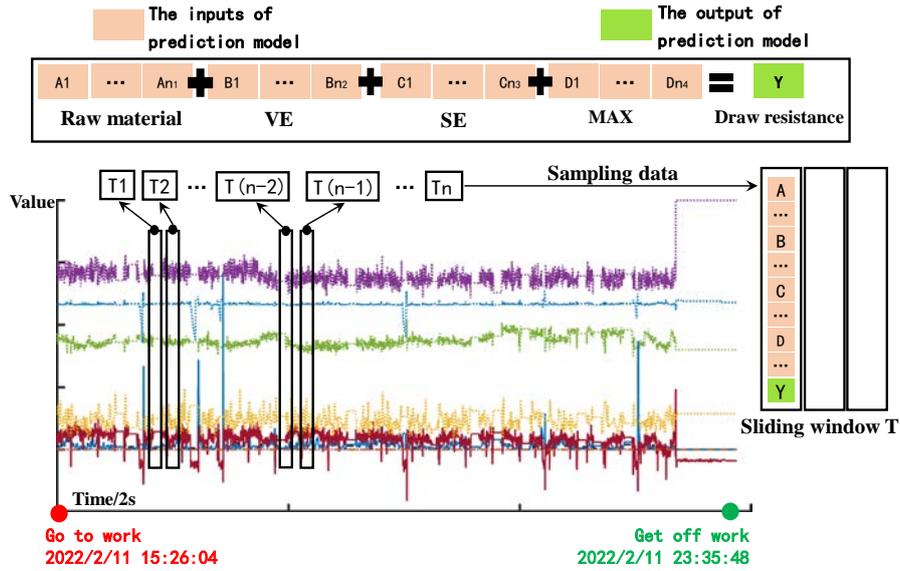

**Figure 14** Sample data

### 4.4 Experimental parameters

The MLFFNN model is a three-layer feed-forward neural network with a hidden layer containing 40 nodes. Five training functions are considered (traingd, trainbr, trainlm, traincgb, trainscg) and the best-performing training function is selected.

For the GMDH model, the number of network layers $L=5$, the maximum number of neurons per layer $N=50$, and the selection pressure $P=0.6$.

For the deep learning model LSTM, the number of hidden layer layers $L=1$, the number of hidden layer dimensions $N=128$, the parameter of dropoutLayer is 0.2, and the maximum number of iterations is 100.

For the ANFIS model, two methods (SC, FCM) are considered to build the model. The core parameters of ANFIS(SC) are Influence radius ($I\_R$) which is taken as 0.5 or 0.6, and the core parameters of ANFIS(FCM) are Number of clusters ($N\_C$) and Partition matrix exponent ($PME$) which are both taken as 2.

For ANFIS-GA, the ANFIS model is optimized using GA (ANFIS here refers to ANFIS-FCM), the number of clusters is 2, in addition, the maximum number of GA iterations, Crossover Percentage, Mutation Percentage, and Mutation Rate of GA are 2000, 0.3, 0.6, and 0.1, respectively. The number of population ($N$) of 10/20/25/50 are considered.

For ANFIS-PSO, the number of clusters of ANFIS is 2, and the PSO parameters are set as follows: the number of populations $N$ is 10, 20, 25 and 50, the inertia weight is 0.9, the personal learning coefficient is 0.9, the global learning coefficient is 1.8, and the maximum number of iterations is 2000.

The experimental statistical indicators are mean square error RMSE, root mean square error MSE, and correlation coefficient R. This experiment focuses on the correlation coefficient R of the test set.

### 4.5 Experimental results

The experimental results are shown in Table 6. Considering only the effect of potential/nonlinear correlation indicators, the test set correlation coefficients are only around 0.5, where the performance of GMDH is relatively good ($MSE=10.4759$, $RMSE=3.5360$, $R=0.5540$).

Considering only the effect of linear indicators, the test set correlation coefficients of ANFIS(FCM), GMDH, LSTM, ANFIS-GA, and ANFIS-PSO models are greater than 0.9, and the best performance is ANFIS-GA.

Linear and non-linear indicators are used together to predict the cigarette draw resistance, and the correlation coefficients of the test set for all models are greater than 0.94, with the best performance of ANFIS-GA ($MSE=1.4357$, $RMSE=1.1982$, $R=0.9684$)

As can be seen in Table 6, although only the direct indicators are considered, the model can perform well. However, the model performs best when both direct and potential indicators are considered.

As shown in Fig. 15, the sudden decrease in the draw resistance, which situation can be accurately predicted by prediction models. The results of correlation analysis can assist the CIS in smoking resistance monitoring and identify substandard cigarettes that are missed by the CIS. Using the probability of substandard cigarettes being sampled as an example, this paper will further highlight the practical value of the prediction method.

A work shift is divided into 100 periods, $j_1 = 100$. One cigarette maker produces about 43200 cigarettes in each period, $j_s \approx 43200$. One shift sampling 200 cigarettes, $Z_1 \approx 200$. Based on the prediction results, the number of substandard cigarettes is 1055, $m_1 = 1055$. Combining with Eq. (16), $P_{old}=0.047$.

To involve the majority of substandard cigarettes in the sampling, 99% of the substandard cigarettes were clearly located to specific production periods in this paper, $m_2 = 0.99 m_1$.

Based on the predicted results, the number of substandard cigarettes is $m_1 = 1055$, then fitting to a normal curve with a standard deviation $\sigma = 1.16$, so $j_2 \approx 6$. One shift sampling 200 cigarettes, $Z_2 = 200$. Combining with Eqs. (22)-(24), $P_{new}=0.558$, the probability of a substandard cigarette being sampled has increased $\Delta P \approx 0.511$. In actual production, five shifts were randomly selected for validation. As the traditional method lacks the characteristics of the distribution of substandard cigarettes, substandard cigarettes cannot be located. However, the prediction method based on correlation analysis proposed in this paper is able to locate substandard cigarettes and increase the probability of substandard cigarettes being sampled. As shown in Table 7, based on the methods in this paper, the substandard cigarettes in the 2022-03-14 Mid, 2022-03-29 Mid, and 2022-03-31 Mid are able to be found, which shows that the correlation analysis-based method for predicting draw resistance performs well, and has a strong practical value as it can locate the production period of substandard cigarettes.

**Table 5** Indicators used in the experiment

| Raw material | VE unit | SE unit | MAX unit |
|---|---|---|---|
| *direct/linear correlation* | | | |
| Long filament rate | Suction rod conveyor vacuum pressure | Cigarette circumference | Ventilating strength |
| Froth filament rate | Suction tape position | Weight average | Ventilating strength of the front channel |
| Short filament rate | The weight of cigarette compaction end | Weight deviation | Ventilation standard deviation of the front channel |
| *potential/nonlinear correlation* | Pressure of air distributor boxes | Short-term standard deviation of weight | Air leakage |
| Tipping paper temperature | Suction tape tension pressure | Weight of cigarettes in sections 1 | Air leakage of the front channel |
| Moisture content of the packing | Target weight | Weight of cigarettes in sections 5 | Air leakage standard deviation of the front channel |
| Filling power | *potential/nonlinear correlation* | The SE belt tension | Relating to cigarette end density of the front channel |
| | Moisture content of the tobacco in the VE | Speed of machine operation | Relating to cigarette end density standard deviation of the front channel |
| | Upper pressure of fluidized bed trough | *potential/nonlinear correlation* | |
| | The pressure of primary separator | The first solder iron temperature | *potential/nonlinear correlation* |
| | Tobacco temperature in tobacco storage | The second solder iron temperature | Rolling plate temperature |
| | | Long-term standard deviation of weight | |
| | | The shell temperature | |
| | | The diode temperature | |
| | | The resonator temperature | |

**Table 6** Experimental results

| Method | Parameters | MSE | | RMSE | | R | |
|---|---|---|---|---|---|---|---|
| | | Train | Test | Train | Test | Train | Test |
| *Input Indexes* | *nonlinear correlation* | | | | | | |
| MLFFNN | rainbr | 16.8604 | 21.0568 | 4.1061 | 4.5888 | 0.2597 | 0.234 |
| ANFIS(SC) | I_R=0.5 | 14.9296 | 15.1718 | 3.8639 | 3.8951 | 0.4466 | 0.4521 |
| ANFIS(FCM) | N_C=2,PME=2 | 14.5184 | 12.8724 | 3.8103 | 3.5878 | 0.4838 | 0.5071 |
| GMDH | N=50,L=2,P=0.5 | 14.7841 | 10.4759 | 3.845 | 3.2366 | 0.4856 | 0.554 |
| LSTM | L=1,N=128,dropout=0.2 | 14.5832 | 12.5036 | 3.8188 | 3.536 | 0.4803 | 0.5273 |
| ANFIS-GA | N=10 | 14.7469 | 10.5974 | 3.8402 | 3.2554 | 0.4876 | 0.5462 |
| ANFIS-PSO | N=20 | 14.4006 | 13.5418 | 3.7948 | 3.6799 | 0.4831 | 0.5094 |
| *Input Indexes* | *linear correlation* | | | | | | |
| MLFFNN | trainlm | 5.5118 | 7.2611 | 2.3477 | 2.6946 | 0.8394 | 0.7917 |
| ANFIS(SC) | I_R=0.6 | 5.4993 | 8.4394 | 2.3451 | 2.9051 | 0.7748 | 0.8362 |
| ANFIS(FCM) | N_C=2,PME=2 | 2.9341 | 3.7448 | 1.7129 | 1.9351 | 0.9157 | 0.9121 |
| GMDH | N=50,L=5,P=0.5 | 2.9119 | 3.4157 | 1.7064 | 1.8482 | 0.9165 | 0.9188 |
| LSTM | L=1,N=128,dropout=0.2 | 2.3967 | 2.4287 | 1.5481 | 1.5587 | 0.9348 | 0.9265 |
| ANFIS-GA | N=50 | 2.3696 | 2.1674 | 1.5394 | 1.4722 | 0.9354 | 0.9351 |
| ANFIS-PSO | N=50 | 2.2085 | 2.5867 | 1.4861 | 1.6083 | 0.9398 | 0.9237 |
| *Input Indexes* | *linear and nonlinear correlation* | | | | | | |
| MLFFNN | trainlm | 1.8242 | 2.0402 | 1.3507 | 1.4284 | 0.9497 | 0.9455 |
| ANFIS(SC) | I_R=0.6 | 1.9874 | 1.929 | 1.4098 | 1.3889 | 0.9447 | 0.9507 |
| ANFIS(FCM) | N_C=2,PME=2 | 1.8016 | 1.8817 | 1.3422 | 1.3718 | 0.9498 | 0.9531 |
| GMDH | N=50,L=7,P=0.5 | 1.829 | 1.7268 | 1.3524 | 1.3141 | 0.9491 | 0.9564 |
| LSTM | L=1,N=128,dropout=0.2 | 1.8409 | 1.478 | 1.3687 | 1.2157 | 0.9484 | 0.9641 |
| ANFIS-GA | N=50 | 1.6853 | 1.4357 | 1.2982 | 1.1982 | 0.9614 | 0.9684 |
| ANFIS-PSO | N=50 | 1.6605 | 1.6811 | 1.2886 | 1.2966 | 0.954 | 0.9571 |

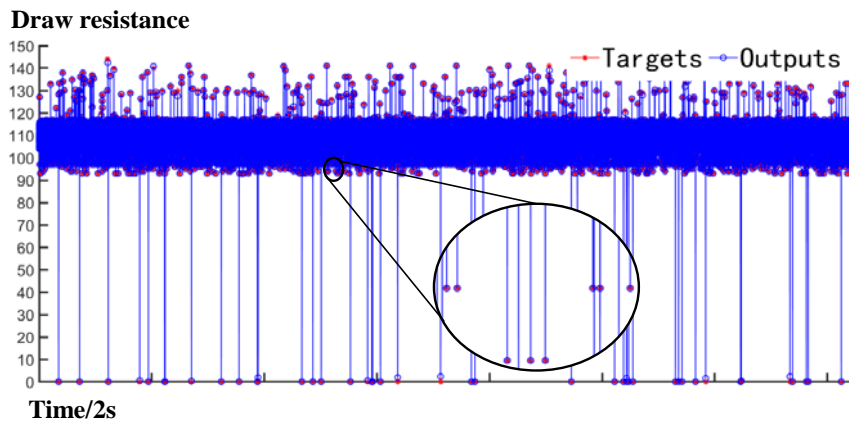

**Figure 15** Prediction results of the cigarette draw resistance

**Table 7** Further highlighting the practical value of predictive models

| Work shift (Morning-Mor, Middle-Mid) | Whether substandard cigarettes are found | |
|---|---|---|
| | old method | our method |
| 2022-03-14Mid | No | Yes |
| 2022-03-29Mor | Yes | Yes |
| 2022-03-29Mid | No | Yes |
| 2022-03-31Mid | No | Yes |
| 2022-04-13Mid | No | No |

Comparing part of the research with this paper, the results are shown in Table 8. For the accuracy of draw resistance monitoring, the CIS, [25], and this paper all achieved 96% accuracy. The CIS is a direct measurement, and [25] only uses a small amount of offline data for modeling and cannot predict online. For the factors used, [26]-[27] considered both linear and non-linear factors, but did not use the expert experience to perform a mechanistic analysis, for example, the effect of temperature on measurement was ignored in [26]. For guidance on follow-up sampling, this paper builds a probabilistic model based on the prediction results of draw resistance, which provides a theoretical basis for sampling and makes anomaly location feasible. In summary, the cigarette draw resistance prediction method proposed in this paper benefits from a rigorous cleaning of the raw data and a comprehensive analysis of the production process.

**Table 8** Comparison results between this paper and other researches

| Method | Accuracy | Offine or Online (time granularity) | Data Cleaning | Linear or nonlinear factors | Sampling is guided |
|---|---|---|---|---|---|
| CIS | >96% | Online | No | / | / |
| Wang[21] | >90% | Offine | No | Linear(7) | No |
| Zhao[23] | >95% | Offine | No | Linear(10) | No |
| Chen[25] | >96% | Online (Not Implemented) | No | Linear(9) | No |
| Chang[24] | >92% | Online (Not Implemented) | No | Linear(5) | No |
| Pang[26] | >95% | Online(1min) | Yes | Linear(30)+nonliear(11) (insufficient) | No |
| Tang[27] | >94% | Online(2s) | Yes | Linear(12)+nonliear(6) (insufficient) | No |
| This paper | >96% | Online(2s) | Yes | Linear(25)+nonliear(14) | Yes |

5. Discussion and the future work

In this paper, to enrich the monitoring method of draw resistance without increasing the hardware cost, the cigarette draw resistance prediction based on correlation analysis is proposed in this paper. The solution contains the pre-analysis of influencing factors based on the mechanism level (raw material, the VE, the SE, and the MAX units), and the indicators are further identified based on linear and non-linear correlation analysis. In addition, machine learning models (MLFFNN, ANFIS, improved ANFIS, GMDH, LSTM) are used to achieve the prediction of the draw resistance (the prediction accuracy is improved to 96.84%, which benefits from the potential factors are additionally considered). Finally, a mathematical model of substandard cigarettes being sampled was innovatively derived, and the practical value of the predictive model is further validated by case studies. In summary, this paper's multi-indicator correlation analysis-based method for predicting cigarette smoking resistance can enrich smoking resistance monitoring methods and provide theoretical support for quality tracing and abnormality location.

For the next step, we will continue to search for more reliable and less time costly methods for monitoring substandard cigarettes, such as building a graph database based on existing indicators and using graph neural networks (GNN) for cigarette quality monitoring.